\documentclass[10pt,a4paper,english,twocolumn, conference]{IEEEtran}
\usepackage[T1]{fontenc}
\usepackage{verbatim}
\usepackage{float}
\usepackage{calc}
\usepackage{textcomp}
\usepackage{mathrsfs}
\usepackage{amsmath}
\usepackage{amsthm}
\usepackage{amssymb}
\usepackage{graphicx}

\makeatletter

\pdfpageheight\paperheight
\pdfpagewidth\paperwidth

\floatstyle{ruled}
\newfloat{algorithm}{tbp}{loa}
\providecommand{\algorithmname}{Algorithm}
\floatname{algorithm}{\protect\algorithmname}

\theoremstyle{plain}
\newtheorem{thm}{\protect\theoremname}
\theoremstyle{plain}
\newtheorem{lem}[thm]{\protect\lemmaname}

\usepackage{subfigure}
\usepackage{epstopdf}
\usepackage{cite}
\usepackage{citesort}

\makeatother

\usepackage{babel}
\providecommand{\lemmaname}{Lemma}
\providecommand{\theoremname}{Theorem}

\begin{document}
% paper title

\title{Ultra-Dense Networks: Is There a Limit to \\
Spatial Spectrum Reuse?% <-this % stops a space
}

\author{{\normalsize{}Ming Ding$^{\ddagger}$, David L$\acute{\textrm{o}}$pez-P$\acute{\textrm{e}}$rez$^{\dagger}$}\emph{\normalsize{},
}{\normalsize{}Guoqiang Mao$^{\nparallel}{}^{\ddagger}$}\emph{\normalsize{},
}{\normalsize{}Zihuai Lin}\textit{\normalsize{}$^{\mathsection}$}\emph{\normalsize{}}\\
\textit{\small{}$^{\ddagger}$Data61, CSIRO, Australia }{\small{}\{Ming.Ding@data61.csiro.au\}}\textit{\small{}}\\
\textit{\small{}$^{\dagger}$Nokia Bell Labs, Ireland }{\small{}\{david.lopez-perez@nokia-bell-labs.com\}}\textit{\small{}}\\
\textit{\small{}$^{\nparallel}$School of Computing and Communication,
University of Technology Sydney, Australia}{\small{}}\\
\textit{\small{}$^{\mathsection}$School of Electrical and Information
Engineering, The University of Sydney, Australia}{\small{}% <-this % stops a space
}}
\maketitle
\begin{abstract}
The aggressive spatial spectrum reuse (SSR) by network densification
using smaller cells has successfully driven the wireless communication
industry onward in the past decades. In our future journey toward
ultra-dense networks (UDNs), a fundamental question needs to be answered.
\emph{Is there a limit to SSR? }In other words, when we deploy thousands
or millions of small cell base stations (BSs) per square kilometer,
is activating all BSs on the same time/frequency resource the best
strategy? In this paper, we present theoretical analyses to answer
such question. In particular, we find that both the signal and interference
powers become bounded in practical UDNs with a non-zero BS-to-UE antenna
height difference and a finite UE density, which leads to a \emph{constant}
capacity scaling law. As a result, there exists an optimal SSR density
that can maximize the network capacity. Hence, the limit to SSR should
be considered in the operation of future UDNs.%
{} 
\end{abstract}

\section{Introduction\label{sec:Introduction}}

\begin{comment}
Placeholder
\end{comment}

From 1950 to 2000, the wireless network capacity has increased around
1$\,$million fold, in which an astounding 2700\texttimes{} gain was
achieved through an aggressive spatial spectrum reuse (SSR) via network
densification using smaller cells~\cite{Webb_survey}. Generally
speaking, SSR means that all the cells in the area of interest simultaneously
reuse the same frequency spectrum. Thus, the wireless network capacity
has the potential to grow linearly as the SSR increases, as each cell
can make an independent and equal contribution to it, given that the
inter-cell interference remains tolerable. The aforementioned 2700\texttimes{}
gain stands as a glorious testimony to the fulfillment of such potential. 

In the first decade of 2000, network densification continued to fuel
the 3rd Generation Partnership Project (3GPP) 4th-generation (4G)
Long Term Evolution (LTE) networks, and is expected to remain as one
of the main forces to drive the 5th-generation (5G) New Radio (NR)
beyond 2020~\cite{Tutor_smallcell}. In particular, the orthogonal
deployment of dense small cell networks (SCNs), in which small cells
and macrocells operate in different frequency bands~\cite{Tutor_smallcell}%
\begin{comment}
 (i.e., the 3GPP SCN Scenario \#2a in~\cite{TR36.872})
\end{comment}
, have gained much momentum in the past years. This is because such
deployment provides a large SSR with easy network management due to
the avoidance of inter-tier interference. 

However, as we walk down the path of network densification, and gradually
enter the realm of ultra-dense networks (UDNs), things start to deviate
from the traditional understanding. In particular, several fundamental
questions arise:
\begin{itemize}
\item The signal power of a typical user equipment (UE) should increase
as a network goes ultra-dense. But is there a limit to such increase
of the signal power?
\item The aggregate interference power of the typical UE should also increase
as a network goes ultra-dense. But is there a limit to such increase
of the aggregate interference power? 
\item Which component will grow faster as a network densifies, the signal
or the aggregate interference power?
\item More importantly, \emph{is there a limit to the SSR?} In other words,
when we deploy thousands or millions of small cell base stations (BSs)
per square kilometer, is activating all BSs on the same time/frequency
resource the best strategy, as we have practiced in the last half
century? Should we explore alternative solutions?%
\begin{comment}
In this paper, we focus on the analysis of these dense SCNs as they
go ultra-dense (UD) in 5G, a.k.a. , and shed new light on their capacity
scaling law. 
\end{comment}
\begin{comment}
Indeed, the orthogonal deployment of dense SCNs within the existing
macrocell networks~\cite{TR36.872}, i.e., small cells and macrocells
operating on different frequency spectrum (Small Cell Scenario \#2a~\cite{TR36.872}),
has been selected as the workhorse for capacity enhancement in the
4th-generation (4G) and the 5th-generation (5G) networks, developed
by the 3rd Generation Partnership Project (3GPP)~\cite{Tutor_smallcell}.
This is due to its large spectrum reuse and its easy management; the
latter one arising from its low interaction with the macrocell tier,
e.g., no inter-tier interference. 
\end{comment}
\end{itemize}
\begin{comment}
Placeholder
\end{comment}

In this paper, we answer this fundamental question via theoretical
analyses. 

\section{Related Work\label{sec:Related-Work}}

\begin{comment}
Placeholder
\end{comment}

Before 2015, the common understanding on UDNs was that the density
of BSs would not affect the per-BS coverage probability performance~\cite{Jeff2011}
in an interference-limited\footnote{In an interference-limited network, the power of each BS is set to
a value much larger than the noise power. } and fully-loaded\footnote{In a fully-loaded network, all BSs are active to generate a full SSR.
Such assumption implies that the UE density is infinite or much larger
than the BS density. According to~\cite{dynOnOff_Huang2012}, the
UE density should be at least 10 times higher than the BS density
to make sure that almost all BSs are active. } wireless network, where the coverage probability is defined as the
probability that the signal-to-interference-plus-noise ratio (SINR)
of a typical UE is above a SINR threshold $\gamma$.%
\begin{comment}
Such phenomenon is referred to as the SINR invariance and is shown
in Fig.~\ref{fig:comp_p_cov_4Gto5G}.
\begin{figure}
\noindent \begin{centering}
\includegraphics[width=8cm,bb = 0 0 200 100, draft, type=eps]{Fig02_PrCov_v4.png}\renewcommand{\figurename}{Fig.}\caption{\label{fig:comp_p_cov_4Gto5G}Theoretical performance comparison of
the coverage probability when the SINR threshold $\gamma=0$\,dB.
Note that all the results are obtained using practical 3GPP channel
models~\cite{TR36.828}, which will be introduced in details later.
Moreover, the BS density regions for the 4G and 5G networks have been
illustrated in the figure, considering that the maximum BS density
of a 4G SCN is in the order of $100\,\textrm{BSs/km}^{2}$~\cite{TR36.872}. }
\par\end{centering}
\vspace{-0.4cm}
\end{figure}
\end{comment}
\begin{comment}
\emph{The SINR invariance} in UDNs predicted by~\cite{Jeff2011}
is shown in Fig.~\ref{fig:comp_p_cov_4Gto5G}.
\end{comment}
{} Such phenomenon is referred to as the SINR invariance. The intuition
of the SINR invariance is that the increase in the aggregate interference
power caused by a denser network would be exactly compensated by the
increase in the signal power due to the reduced distance between transmitters
and receivers~\cite{Jeff2011}. Consequently, the network capacity
should scale \emph{linearly} as the BS density increases in a fully-loaded
UDN. Such conclusion, however, was obtained with considerable simplifications
on network conditions and propagation environment. 

Recently, a few noteworthy studies have followed and revisited the
network performance analysis of UDNs using more practical assumptions~\cite{related_work_Jeff,Related_work_Health,our_work_TWC2016,Renzo2016intensityMatch,Ding2016ASECrash,Ding2016TWC_IMC},
such as
\begin{itemize}
\item a general multi-piece path loss model with probabilistic line-of-sight
(LoS) and non-LoS (NLoS) transmissions,
\item a non-zero BS-to-UE antenna height difference $L$, and
\item a non-fully-loaded network with a finite UE density $\rho$.
\end{itemize}
\begin{comment}
Placeholder
\end{comment}

The inclusion of these more realistic assumptions significantly changed
the previous conclusion on the SINR invariance~\cite{Jeff2011},
indicating that the coverage probability performance of UDNs is \emph{neither
a convex nor a concave function} with respect to the BS density. In
particular, two seemingly contradictory performance behaviors can
be observed in~\cite{Ding2016ASECrash} and~\cite{Ding2016TWC_IMC},
both considering a general multi-piece path loss model recommended
by the 3GPP.

First, if we consider a practical non-zero BS-to-UE antenna height
difference $L$, then the coverage probability is shown to crash as
the BS density increases in a fully-loaded UDN. This is caused by
a severe \emph{SINR decrease} in UDNs~\cite{Ding2016ASECrash}%
\begin{comment}
, as exhibited in Fig.~\ref{fig:comp_p_cov_4Gto5G} with $L=8.5$\,m,
where the SCN BS antenna height and the UE antenna height are assumed
to be 10\,m and 1.5\,m, respectively~\cite{TR36.828}
\end{comment}
. The intuition of such \emph{SINR decrease} is that the signal power
becomes \emph{bounded} in UDNs due to the lower-bound on the BS-to-UE
distance, as a UE cannot be closer than $L$ to its serving BS. 

Second, if we consider a practical finite UE density $\rho$, then
the coverage probability is shown to take off as the BS density increases.
This is caused by a soaring \emph{SINR increase} in UDNs~\cite{Ding2016TWC_IMC}%
\begin{comment}
, as exhibited in Fig.~\ref{fig:comp_p_cov_4Gto5G} with a UE density
of $\rho=300\thinspace\textrm{UEs/km}^{2}$, where this is a typical
UE density in populated scenarios in 5G~\cite{Tutor_smallcell}
\end{comment}
. The intuition of such \emph{SINR increase} is that the aggregate
interference power becomes \emph{bounded} in UDNs due to the partial
activation of a finite density of BSs to serve a finite density of
UEs. In more detail, a large number of BSs can switch off their transmission
modules in UDNs, entering into idle mode, if there is no active UE
within their coverage areas. As a result, the number of interfering
BSs and also the SSR are limited by the finite number of UEs. 

Considering that the above two seemingly contradictory performance
behaviors (i.e., SINR\emph{ decrease} and \emph{increase}) manifest
themselves in UDNs, it is of great interest to investigate their trade-offs.
\emph{Which one prevails in UDN}s? Such study will eventually reveal
the answer to the fundamental question: \emph{Is there a limit to
the SSR?} Our short answer is YES.%
\begin{comment}
, while the signal power is unbounded. (Actually it is bounded by
a very large value of 1) 
\end{comment}
\begin{comment}
an intriguing question rises: \emph{What occurs in reality when both
effects are considered?} In this paper, we answer this fundamental
question by theoretical analyses. In particular, 
\end{comment}
\textbf{}%
\begin{comment}
\textbf{We present and prove the existence of a new SINR invariance
in UDNs under practical assumptions.}

\textbf{Then, we discover and prove a new capacity scaling law in
UDNs, which is a constant scaling law. }
\end{comment}
\begin{comment}
The rest of this paper is structured as follows. Section~\ref{sec:System-Model}
describes the network scenario and the wireless system model considered
in this paper. Section~\ref{sec:Main-Results} presents our theoretical
results on the coverage probability and the area spectral efficiency
(ASE), followed by our discoveries of \emph{a new SINR invariance}
and \emph{a new capacity scaling law} in UDNs. The numerical results
are discussed in Section~\ref{sec:Simulation-and-Discussion}, with
remarks shedding light on the significance of the discovered capacity
scaling law in UDNs. The conclusions are drawn in Section~\ref{sec:Conclusion}.
\end{comment}

\section{Network Scenario and System Model\label{sec:System-Model}}

\begin{comment}
Placeholder
\end{comment}

In this section, we present the network scenario and the wireless
system model considered in this paper. 

\subsection{Network Scenario\label{subsec:Network-Scenario}}

\begin{comment}
Placeholder
\end{comment}

We consider a downlink (DL) cellular network with BSs deployed on
a plane according to a homogeneous Poisson point process (HPPP) $\Phi$
with a density of $\lambda$ $\textrm{BSs/km}^{2}$. Active DL UEs
are also Poisson distributed in the considered network with a density
of $\rho$ $\textrm{UEs/km}^{2}$. Here, we only consider active UEs
in the network because non-active UEs do not trigger any data transmission%
\begin{comment}
, and thus they can be safely ignored in our analysis
\end{comment}
\begin{comment}
Note that the total UE number in cellular networks should be much
higher than the number of the active UEs, but at a certain time slot
and on a certain frequency band, the active UEs with data traffic
demands are not too many. 
\end{comment}
, the typical density of which is around $\rho=300\thinspace\textrm{UEs/km}^{2}$~\cite{Tutor_smallcell}.

In practice, a BS will enter an idle mode if there is no UE connected
to it, which reduces the interference to neighboring UEs as well as
the energy consumption of the network. The set of active BSs is thus
depending on the user association strategy (UAS). In this paper, we
assume a practical UAS as in~\cite{our_work_TWC2016}, where each
UE is connected to the BS having the maximum average received signal
strength, which will be formally presented in Subsection~\ref{subsec:Wireless-System-Model}.%
\begin{comment}
In practice, a BS will enter into idle mode, if there is no UE connected
to it, which reduces the interference to UEs in neighboring BSs as
well as the energy consumption of the network. 
\end{comment}
{} Since UEs are randomly and uniformly distributed in the network,
the active BSs should follow another HPPP distribution $\tilde{\Phi}$,
the density of which is $\tilde{\lambda}$ $\textrm{BSs/km}^{2}$
\cite{dynOnOff_Huang2012}. Such $\tilde{\lambda}$ also characterizes
the SSR because only active BSs use the frequency spectrum. Moreover,
note that $\tilde{\lambda}\leq\lambda$ and $\tilde{\lambda}\leq\rho$,
since one UE is served by at most one BS, and that a larger $\rho$
results in a larger $\tilde{\lambda}$. From~\cite{dynOnOff_Huang2012},
$\tilde{\lambda}$ can be calculated as%
\begin{comment}
 it was shown that the formula proposed in~\cite{dynOnOff_Huang2012}
to calculate $\tilde{\lambda}$ is accurate for UDNs, which is given
by
\end{comment}
\begin{equation}
\tilde{\lambda}=\lambda\left[1-\frac{1}{\left(1+\frac{\rho}{q\lambda}\right)^{q}}\right],\label{eq:lambda_tilde_Huang}
\end{equation}
where an empirical value of 3.5 was suggested for $q$ in~\cite{dynOnOff_Huang2012}\footnote{Note that according to~\cite{Ding2016TWC_IMC}, $q$ should also
depend on the path loss model, which will be presented in Subsection~\ref{subsec:Wireless-System-Model}.
Having said that, \cite{Ding2016TWC_IMC} also showed that (\ref{eq:lambda_tilde_Huang})
is generally very accurate to characterize $\tilde{\lambda}$ for
dense and ultra-dense networks.}. 

\subsection{Wireless System Model\label{subsec:Wireless-System-Model}}

\begin{comment}
Placeholder
\end{comment}

The two-dimensional (2D) distance between a BS and a UE is denoted
by $r$. Moreover, the absolute antenna height difference between
a BS and a UE is denoted by $L$. Thus, the 3D distance between a
BS and a UE can be expressed as
\begin{equation}
w=\sqrt{r^{2}+L^{2}}.\label{eq:actual_dis_BS2UE}
\end{equation}
Note that the value of $L$ is in the order of several meters~\cite{TR36.828}.

Following~\cite{our_work_TWC2016}, we adopt a general path loss
model, where the path loss $\zeta\left(w\right)$ is a multi-piece
function of $w$ written as%
\begin{comment}
\begin{singlespace}
\noindent 
\[
\zeta\left(w\right)=\begin{cases}
\zeta_{1}\left(w\right)=\begin{cases}
\begin{array}{l}
\zeta_{1}^{\textrm{L}}\left(w\right),\\
\zeta_{1}^{\textrm{NL}}\left(w\right),
\end{array} & \hspace{-0.3cm}\begin{array}{l}
\textrm{with probability }\textrm{Pr}_{1}^{\textrm{L}}\left(w\right)\\
\textrm{with probability }\left(1-\textrm{Pr}_{1}^{\textrm{L}}\left(w\right)\right)
\end{array}\end{cases}\hspace{-0.3cm}, & \hspace{-0.3cm}\textrm{when }0\leq w\leq d_{1}\\
\zeta_{2}\left(w\right)=\begin{cases}
\begin{array}{l}
\zeta_{2}^{\textrm{L}}\left(w\right),\\
\zeta_{2}^{\textrm{NL}}\left(w\right),
\end{array} & \hspace{-0.3cm}\begin{array}{l}
\textrm{with probability }\textrm{Pr}_{2}^{\textrm{L}}\left(w\right)\\
\textrm{with probability }\left(1-\textrm{Pr}_{2}^{\textrm{L}}\left(w\right)\right)
\end{array}\end{cases}\hspace{-0.3cm}, & \hspace{-0.3cm}\textrm{when }d_{1}<w\leq d_{2}\\
\vdots & \vdots\\
\zeta_{N}\left(w\right)=\begin{cases}
\begin{array}{l}
\zeta_{N}^{\textrm{L}}\left(w\right),\\
\zeta_{N}^{\textrm{NL}}\left(w\right),
\end{array} & \hspace{-0.3cm}\begin{array}{l}
\textrm{with probability }\textrm{Pr}_{N}^{\textrm{L}}\left(w\right)\\
\textrm{with probability }\left(1-\textrm{Pr}_{N}^{\textrm{L}}\left(w\right)\right)
\end{array}\end{cases}\hspace{-0.3cm}, & \hspace{-0.3cm}\textrm{when }w>d_{N-1}
\end{cases}.
\]
\end{singlespace}
\end{comment}
\begin{equation}
\zeta\left(w\right)=\begin{cases}
\zeta_{1}\left(w\right), & \textrm{when }L\leq w\leq d_{1}\\
\zeta_{2}\left(w\right), & \textrm{when }d_{1}<w\leq d_{2}\\
\vdots & \vdots\\
\zeta_{N}\left(w\right), & \textrm{when }w>d_{N-1}
\end{cases},\label{eq:prop_PL_model}
\end{equation}
where each piece $\zeta_{n}\left(w\right),n\in\left\{ 1,2,\ldots,N\right\} $
is modeled as
\begin{equation}
\zeta_{n}\left(w\right)\hspace{-0.1cm}=\hspace{-0.1cm}\begin{cases}
\hspace{-0.2cm}\begin{array}{l}
\zeta_{n}^{{\rm {L}}}\left(w\right)=A_{n}^{{\rm {L}}}w^{-\alpha_{n}^{{\rm {L}}}},\\
\zeta_{n}^{{\rm {NL}}}\left(w\right)=A_{n}^{{\rm {NL}}}w^{-\alpha_{n}^{{\rm {NL}}}},
\end{array} & \hspace{-0.2cm}\hspace{-0.3cm}\begin{array}{l}
\textrm{LoS:}~\textrm{Pr}_{n}^{{\rm {L}}}\left(w\right)\\
\textrm{NLoS:}~1-\textrm{Pr}_{n}^{{\rm {L}}}\left(w\right)
\end{array}\hspace{-0.1cm}\hspace{-0.1cm},\hspace{-0.1cm}\hspace{-0.1cm}\end{cases}\label{eq:PL_BS2UE}
\end{equation}
where%
\begin{comment}
\noindent In practice, $A_{n}^{{\rm {L}}}$, $A_{n}^{{\rm {NL}}}$,
$\alpha_{n}^{{\rm {L}}}$ and $\alpha_{n}^{{\rm {NL}}}$ are constants
obtainable from field tests~\cite{TR36.828}. 
\end{comment}

\begin{itemize}
\item $\zeta_{n}^{{\rm {L}}}\left(w\right)$ and $\zeta_{n}^{{\rm {NL}}}\left(w\right),n\in\left\{ 1,2,\ldots,N\right\} $
are the $n$-th piece path loss functions for the LoS and the NLoS
cases, respectively,
\item $A_{n}^{{\rm {L}}}$ and $A_{n}^{{\rm {NL}}}$ are the path losses
at a reference 3D distance $w=1$ for the LoS and the NLoS cases,
respectively,
\item $\alpha_{n}^{{\rm {L}}}$ and $\alpha_{n}^{{\rm {NL}}}$ are the path
loss exponents for the LoS and the NLoS cases, respectively.
\end{itemize}
Moreover, $\textrm{Pr}_{n}^{{\rm {L}}}\left(w\right)$ is the $n$-th
piece LoS probability function that a transmitter and a receiver separated
by a 3D distance $w$ has an LoS path, which is assumed to be \emph{a
monotonically decreasing function} with respect to $w$. Existing
measurement studies have confirmed this assumption~\cite{TR36.828}.%
\begin{comment}
For convenience, $\left\{ \zeta_{n}^{\textrm{L}}\left(r\right)\right\} $
and $\left\{ \zeta_{n}^{\textrm{NL}}\left(r\right)\right\} $ are
further stacked into piece-wise functions written as
\begin{equation}
\zeta^{Path}\left(r\right)=\begin{cases}
\zeta_{1}^{Path}\left(r\right), & \textrm{when }0\leq r\leq d_{1}\\
\zeta_{2}^{Path}\left(r\right),\hspace{-0.3cm} & \textrm{when }d_{1}<r\leq d_{2}\\
\vdots & \vdots\\
\zeta_{N}^{Path}\left(r\right), & \textrm{when }r>d_{N-1}
\end{cases},\label{eq:general_PL_func}
\end{equation}
where the string variable $Path$ takes the value of ``L'' and ``NL''
for the LoS and the NLoS cases, respectively. 
\end{comment}
\begin{comment}
For convenience, $\left\{ \textrm{Pr}_{n}^{\textrm{L}}\left(r\right)\right\} $
is stacked into a piece-wise function as
\begin{equation}
\textrm{Pr}^{\textrm{L}}\left(r\right)=\begin{cases}
\textrm{Pr}_{1}^{\textrm{L}}\left(r\right), & \textrm{when }0\leq r\leq d_{1}\\
\textrm{Pr}_{2}^{\textrm{L}}\left(r\right),\hspace{-0.3cm} & \textrm{when }d_{1}<r\leq d_{2}\\
\vdots & \vdots\\
\textrm{Pr}_{N}^{\textrm{L}}\left(r\right), & \textrm{when }r>d_{N-1}
\end{cases}.\label{eq:general_LoS_Pr}
\end{equation}
\end{comment}

As a special case to show our numerical results in the simulation
section, we consider a practical two-piece path loss function and
a two-piece exponential LoS probability function, defined by the 3GPP~\cite{TR36.828}.
Specifically, we have $N=2$, $\zeta_{1}^{{\rm {L}}}\left(w\right)=\zeta_{2}^{{\rm {L}}}\left(w\right)=A^{{\rm {L}}}w^{-\alpha^{{\rm {L}}}}$,
$\zeta_{1}^{{\rm {NL}}}\left(w\right)=\zeta_{2}^{{\rm {NL}}}\left(w\right)=A^{{\rm {NL}}}w^{-\alpha^{{\rm {NL}}}}$,
$\textrm{Pr}_{1}^{{\rm {L}}}\left(w\right)=1-5\exp\left(-R_{1}/w\right)$,
and $\textrm{Pr}_{2}^{{\rm {L}}}\left(w\right)=5\exp\left(-w/R_{2}\right)$,
where $R_{1}=156$\ m, $R_{2}=30$\ m, and $d_{1}=\frac{R_{1}}{\ln10}=67.75$\ m~\cite{TR36.828}.
For clarity, this path loss case is referred to as \textbf{the 3GPP
Case} hereafter.%
\begin{comment}
Note that the 3GPP Case has been used to generate the results in Fig.~\ref{fig:comp_p_cov_4Gto5G}
of Section~\ref{sec:Introduction}.
\end{comment}
{} 

As discussed before, we assume a practical user association strategy
(UAS), in which each UE is connected to the BS giving the maximum
average received signal strength (i.e., with the largest $\zeta\left(w\right)$)~\cite{Related_work_Health,our_work_TWC2016}.%
{} Finally, we assume that each BS's transmission power has a constant
value $P$, each BS/UE is equipped with an isotropic antenna, and
the multi-path fading between a BS and a UE is modeled as independently
identical distributed (i.i.d.) Rayleigh fading~\cite{related_work_Jeff,Related_work_Health,our_work_TWC2016}.

\subsection{More Network Assumptions in Future Work}

\begin{comment}
Placeholder
\end{comment}

Regarding other assumptions, it is important to note that it has been
shown in~\cite{Ding2017varFactors} through simulation that the analyses
of the following factors/models are not urgent, as they do not change
the qualitative conclusions of this type of performance analysis in
UDNs: 
\begin{itemize}
\item \emph{A deterministic non-Poisson distributed BS/UE density. }
\item \emph{A BS density dependent transmission power.}
\item \emph{A more accurate multi-path modeling with Rician fading. }
\item \emph{An additional modeling of correlated shadow fading. }
\end{itemize}
Thus, we will focus on presenting our most fundamental results in
this paper, and show the minor impacts of the above factors/models
in the journal version of this work.

\section{Main Result\label{sec:Main-Results}}

\begin{comment}
Placeholder
\end{comment}

In this section, we study the coverage probability performance and
the network capacity in terms of the area spectral efficiency (ASE)
of a typical UE located at the origin $o$. 

\subsection{The Coverage Probability\label{subsec:The-Coverage-Probability}}

\begin{comment}
Placeholder
\end{comment}

First, we investigate the coverage probability that the SINR of the
typical UE at the origin $o$ is above a threshold $\gamma$:
\begin{equation}
p^{{\rm {cov}}}\left(\lambda,\rho,\gamma\right)=\textrm{Pr}\left[\mathrm{SINR}>\gamma\right],\label{eq:Coverage_Prob_def}
\end{equation}
where the SINR is computed by
\begin{equation}
\mathrm{SINR}=\frac{P\zeta\left(w\right)h}{I_{{\rm {agg}}}+P_{{\rm {N}}}},\label{eq:SINR}
\end{equation}
where $h$ is the channel gain, which is modeled as an exponentially
distributed random variable (RV) with a mean of one due to our consideration
of Rayleigh fading, presented in Subsection~\ref{subsec:Wireless-System-Model},
$P$ and $P_{{\rm {N}}}$ are the BS transmission power and the additive
white Gaussian noise (AWGN) power at each UE, respectively, and $I_{{\rm {agg}}}$
is the aggregate interference given by
\begin{equation}
I_{{\rm {agg}}}=\sum_{i:\,b_{i}\in\tilde{\Phi}\setminus b_{o}}P\beta_{i}g_{i},\label{eq:cumulative_interference}
\end{equation}
where $b_{o}$ is the BS serving the typical UE, and $b_{i}$, $\beta_{i}$
and $g_{i}$ are the $i$-th interfering BS, the path loss from $b_{i}$
to the typical UE and the multi-path fading channel gain associated
with such link (also exponentially distributed RVs), respectively.
Note that,%
\begin{comment}
when all BSs are assumed to be active, the set of all BSs $\Phi$
should be used in the expression of $I_{\textrm{agg}}$~\cite{related_work_Jeff,Related_work_Health,our_work_TWC2016}.
Here, 
\end{comment}
{} in (\ref{eq:cumulative_interference}), only the BSs in $\tilde{\Phi}\setminus b_{o}$
inject effective interference into the network, where $\tilde{\Phi}$
denotes the set of the active BSs. In other words, the BSs in idle
mode are not taken into account in the computation of $I_{{\rm {agg}}}$. 

Based on the general path loss model in (\ref{eq:prop_PL_model})
and the adopted UAS, in Theorem~\ref{thm:p_cov_limit_UAS1}, we present
our main result on the asymptotic performance of $p^{{\rm {cov}}}\left(\lambda,\rho,\gamma\right)$
in UDNs, i.e., $\underset{\lambda\rightarrow+\infty}{\lim}p^{\textrm{cov}}\left(\lambda,\rho,\gamma\right)$.{\small{}}
\begin{algorithm*}
\begin{thm}
\label{thm:p_cov_limit_UAS1}Considering the general path loss model
in (\ref{eq:prop_PL_model}) and the adopted UAS, $\underset{\lambda\rightarrow+\infty}{\lim}p^{\textrm{cov}}\left(\lambda,\rho,\gamma\right)$
can be derived as
\begin{equation}
\underset{\lambda\rightarrow+\infty}{\lim}p^{\textrm{cov}}\left(\lambda,\rho,\gamma\right)=\underset{\lambda\rightarrow+\infty}{\lim}{\rm {Pr}}\hspace{-0.1cm}\left[\frac{P\zeta_{1}^{{\rm {L}}}\left(L\right)h}{I_{{\rm {agg}}}+P_{{\rm {N}}}}\hspace{-0.1cm}>\hspace{-0.1cm}\gamma\right]=\exp\left(-\frac{P_{{\rm {N}}}\gamma}{P\zeta_{1}^{{\rm {L}}}\left(L\right)}\right)\underset{\lambda\rightarrow+\infty}{\lim}\mathscr{L}_{I_{{\rm {agg}}}}^{{\rm {L}}}\left(\frac{\gamma}{P\zeta_{1}^{{\rm {L}}}\left(L\right)}\right),\label{eq:Theorem_1_p_cov_limit}
\end{equation}
where $\underset{\lambda\rightarrow+\infty}{\lim}\mathscr{L}_{I_{{\rm {agg}}}}^{{\rm {L}}}\left(s\right)$
with $s=\frac{\gamma}{P\zeta_{1}^{{\rm {L}}}\left(L\right)}$ is given
by
\begin{equation}
\underset{\lambda\rightarrow+\infty}{\lim}\mathscr{L}_{I_{{\rm {agg}}}}^{{\rm {L}}}\hspace{-0.1cm}\left(s\right)=\exp\hspace{-0.1cm}\left(\hspace{-0.1cm}-2\pi\rho\hspace{-0.1cm}\int_{0}^{+\infty}\hspace{-0.1cm}\hspace{-0.1cm}\hspace{-0.1cm}\hspace{-0.1cm}\frac{{\rm {Pr}}^{{\rm {L}}}\left(\sqrt{u^{2}+L^{2}}\right)u}{1\hspace{-0.1cm}+\hspace{-0.1cm}\left(sP\zeta^{{\rm {L}}}\left(\sqrt{u^{2}+L^{2}}\right)\right)^{-1}}du\hspace{-0.1cm}\right)\exp\hspace{-0.1cm}\left(\hspace{-0.1cm}-2\pi\rho\hspace{-0.1cm}\int_{0}^{+\infty}\hspace{-0.1cm}\hspace{-0.1cm}\hspace{-0.1cm}\hspace{-0.1cm}\frac{\left[1\hspace{-0.1cm}-\hspace{-0.1cm}{\rm {Pr}}^{{\rm {L}}}\left(\sqrt{u^{2}+L^{2}}\right)\right]u}{1\hspace{-0.1cm}+\hspace{-0.1cm}\left(sP\zeta^{{\rm {NL}}}\left(\sqrt{u^{2}+L^{2}}\right)\right)^{-1}}du\hspace{-0.1cm}\right)\hspace{-0.1cm}.\hspace{-0.1cm}\label{eq:laplace_term_LoS_UAS1_general_seg_thm}
\end{equation}
\end{thm}
\begin{IEEEproof}
See Appendix~A.
\end{IEEEproof}
\end{algorithm*}
\vspace{0.3cm}

From Theorem~\ref{thm:p_cov_limit_UAS1}, we propose a new SINR invariance
law in Theorem~\ref{thm:SINR-invariance-law}.
\begin{thm}
\label{thm:SINR-invariance-law}A new SINR invariance law: If $L>0$
and $\rho<+\infty$, then $\underset{\lambda\rightarrow+\infty}{\lim}p^{\textrm{cov}}\left(\lambda,\rho,\gamma\right)$
becomes a constant that is independent of $\lambda$ in UDNs%
\begin{comment}
, and such constant is determined by $L$, $\rho$ and $\gamma$ as
shown in (\ref{eq:Theorem_1_p_cov_limit}) and (\ref{eq:laplace_term_LoS_UAS1_general_seg_thm})
of Theorem~\ref{thm:p_cov_limit_UAS1}
\end{comment}
. 
\end{thm}
\begin{IEEEproof}
See Appendix~B.
\end{IEEEproof}
\begin{comment}
Placeholder
\end{comment}
\vspace{0.3cm}

Theorem~\ref{thm:SINR-invariance-law} indicates that%
\begin{comment}
in a practical SCN with $L>0$ and $\rho<+\infty$, the coverage probability
$p^{{\rm {cov}}}\left(\lambda,\gamma\right)$ shall approach a constant
in UDNs. In other words, 
\end{comment}
\emph{ (i) }the \emph{SINR decrease} effect due to the non-zero BS-to-UE
antenna height difference $L$ and \emph{(ii) }the \emph{SINR increase}
due to the finite UE density $\rho$ and the BS idle mode capability
counter-balance each other in practical UDNs with $L>0$ and $\rho<+\infty$%
\begin{comment}
, as shown in Fig.~\ref{fig:comp_p_cov_4Gto5G}
\end{comment}
. Note that here the study on $\left\{ L,\rho\right\} $ is finally
complete because:
\begin{itemize}
\item The case of $L=0$ and $\rho=+\infty$ has been studied in~\cite{related_work_Jeff,Related_work_Health,our_work_TWC2016},
showing that $\underset{\lambda\rightarrow+\infty}{\lim}p^{\textrm{cov}}\left(\lambda,\rho,\gamma\right)$
is a function of $\alpha_{n}^{{\rm {L}}}$.
\item The case of $L>0$ and $\rho=+\infty$ has been studied in~\cite{Ding2016ASECrash},
showing that $\underset{\lambda\rightarrow+\infty}{\lim}p^{\textrm{cov}}\left(\lambda,\rho,\gamma\right)=0$%
\begin{comment}
 as illustrated in Fig.~\ref{fig:comp_p_cov_4Gto5G}
\end{comment}
.
\item The case of $L=0$ and $\rho<+\infty$ has been studied in~\cite{Ding2016TWC_IMC},
showing that $\underset{\lambda\rightarrow+\infty}{\lim}p^{\textrm{cov}}\left(\lambda,\rho,\gamma\right)=1$%
\begin{comment}
as illustrated in Fig.~\ref{fig:comp_p_cov_4Gto5G}
\end{comment}
.
\item The case of $L>0$ and $\rho<+\infty$ is characterized by Theorem~\ref{thm:SINR-invariance-law},
which reflects the most practical SCN deployment among the above cases. 
\end{itemize}
\begin{comment}
Placeholder
\end{comment}

From Theorem~\ref{thm:SINR-invariance-law}, it is trivial to show
that for a given $\left\{ L,\rho\right\} $, $\underset{\lambda\rightarrow+\infty}{\lim}p^{\textrm{cov}}\left(\lambda,\rho,\gamma\right)$
decreases as $\gamma$ increases. This is because a higher SINR requirement
naturally leads to a lower coverage probability. Thus, in Lemmas~\ref{lem:Pcov_limit_changes_with_L}
and~\ref{lem:Pcov_limit_changes_with_rou}, we only address how $\underset{\lambda\rightarrow+\infty}{\lim}p^{\textrm{cov}}\left(\lambda,\rho,\gamma\right)$
varies with $L$ and $\rho$, respectively.%
\begin{comment}
However, it is not obvious how $\underset{\lambda\rightarrow+\infty}{\lim}p^{\textrm{cov}}\left(\lambda,\gamma\right)$
varies with $L$ and $\rho$, which will be addressed in Lemmas~\ref{lem:Pcov_limit_changes_with_L}
and~\ref{lem:Pcov_limit_changes_with_rou}, respectively. 
\end{comment}

\begin{lem}
\label{lem:Pcov_limit_changes_with_L}For a given $\left\{ \rho,\gamma\right\} $,
$\underset{\lambda\rightarrow+\infty}{\lim}p^{\textrm{cov}}\left(\lambda,\rho,\gamma\right)$
decreases as $L$ increases. 
\end{lem}
\begin{IEEEproof}
See Appendix~C.
\end{IEEEproof}
\begin{lem}
\label{lem:Pcov_limit_changes_with_rou}For a given $\left\{ L,\gamma\right\} $,
$\underset{\lambda\rightarrow+\infty}{\lim}p^{\textrm{cov}}\left(\lambda,\rho,\gamma\right)$
decreases as $\rho$ increases, according to a power law with respect
to $\rho$. More specifically, we have 
\begin{equation}
\underset{\lambda\rightarrow+\infty}{\lim}p^{\textrm{cov}}\left(\lambda,\rho,\gamma\right)=c\left(\gamma\right)g^{\rho}\left(\gamma\right),\label{eq:decomp_c_func_g_func}
\end{equation}
where $c\left(\gamma\right)$ and $g\left(\gamma\right)$ are expressed
as
\begin{equation}
c\left(\gamma\right)=\exp\left(-\frac{P_{{\rm {N}}}\gamma}{P\zeta_{1}^{{\rm {L}}}\left(L\right)}\right),\label{eq:c_func}
\end{equation}
and
\begin{eqnarray}
\hspace{-0.1cm}\hspace{-0.1cm}\hspace{-0.1cm}\hspace{-0.1cm}\hspace{-0.1cm}\hspace{-0.1cm}g\left(\gamma\right)\hspace{-0.1cm}\hspace{-0.1cm}\hspace{-0.1cm} & = & \hspace{-0.1cm}\hspace{-0.1cm}\hspace{-0.1cm}\exp\hspace{-0.1cm}\left(\hspace{-0.1cm}-2\pi\int_{0}^{+\infty}\hspace{-0.1cm}\hspace{-0.1cm}\hspace{-0.1cm}\hspace{-0.1cm}\frac{{\rm {Pr}}^{{\rm {L}}}\left(\sqrt{u^{2}+L^{2}}\right)u}{1\hspace{-0.1cm}+\hspace{-0.1cm}\left(sP\zeta^{{\rm {L}}}\left(\sqrt{u^{2}+L^{2}}\right)\right)^{-1}}du\hspace{-0.1cm}\right)\nonumber \\
\hspace{-0.1cm}\hspace{-0.1cm}\hspace{-0.1cm} &  & \hspace{-0.1cm}\hspace{-0.1cm}\hspace{-0.1cm}\times\exp\hspace{-0.1cm}\left(\hspace{-0.1cm}-2\pi\int_{0}^{+\infty}\hspace{-0.1cm}\hspace{-0.1cm}\hspace{-0.1cm}\hspace{-0.1cm}\frac{\left[1\hspace{-0.1cm}-\hspace{-0.1cm}{\rm {Pr}}^{{\rm {L}}}\left(\sqrt{u^{2}+L^{2}}\right)\right]u}{1\hspace{-0.1cm}+\hspace{-0.1cm}\left(sP\zeta^{{\rm {NL}}}\left(\sqrt{u^{2}+L^{2}}\right)\right)^{-1}}du\hspace{-0.1cm}\right)\hspace{-0.1cm},\label{eq:laplace_term_LoS_UAS1_g_func}
\end{eqnarray}
where $s=\frac{\gamma}{P\zeta_{1}^{{\rm {L}}}\left(L\right)}$. 
\end{lem}
\begin{IEEEproof}
See Appendix~D.
\end{IEEEproof}
\begin{comment}
Placeholder
\end{comment}

The intuitions of Lemmas~\ref{lem:Pcov_limit_changes_with_L} and~\ref{lem:Pcov_limit_changes_with_rou}
are explained as follows,
\begin{itemize}
\item The signal power becomes bounded in UDNs due to the lower-bound on
the BS-to-UE distance, as a UE cannot be closer than $L$ to a BS.
Moreover, a larger $L$ implies a tighter bound on the signal power,
which leads to the decrease of $\underset{\lambda\rightarrow+\infty}{\lim}p^{\textrm{cov}}\left(\lambda,\rho,\gamma\right)$,
as shown in Lemma~\ref{lem:Pcov_limit_changes_with_L}. 
\item The aggregate interference power becomes bounded in UDNs due to the
activation of a finite density of BSs (i.e., $\tilde{\lambda}$) to
serve a finite density of UEs (i.e., $\rho$). Moreover, a larger
$\rho$ results in a larger $\tilde{\lambda}$, relaxing the bound
on the aggregate interference power, which leads to the decrease of
$\underset{\lambda\rightarrow+\infty}{\lim}p^{\textrm{cov}}\left(\lambda,\rho,\gamma\right)$,
as shown in Lemma~\ref{lem:Pcov_limit_changes_with_rou}. Such decrease
follows a power law with respect to $\rho$, because an HPPP distribution
of UEs with $\rho\thinspace\textrm{UEs/km}^{2}$ can be decomposed
into $\rho$ independent HPPP ones with $1\thinspace\textrm{UEs/km}^{2}$
each, and the coverage criterion (\ref{eq:Coverage_Prob_def}) should
be satisfied for each one of these HPPP distributions. This yields
a power law with respect to $\rho$.
\end{itemize}

\subsection{The Area Spectral Efficiency\label{subsec:The-Area-Spectral}}

\begin{comment}
Placeholder
\end{comment}

Next, we investigate the network capacity performance in terms of
the area spectral efficiency (ASE) in $\textrm{bps/Hz/km}^{2}$, which
is defined as~\cite{our_work_TWC2016}
\begin{equation}
A^{{\rm {ASE}}}\left(\lambda,\rho,\gamma_{0}\right)=\tilde{\lambda}\int_{\gamma_{0}}^{+\infty}\log_{2}\left(1+\gamma\right)f_{\mathit{\Gamma}}\left(\lambda,\rho,\gamma\right)d\gamma,\label{eq:ASE_def}
\end{equation}
where $\gamma_{0}$ is the minimum working SINR in a practical SCN,
and $f_{\mathit{\Gamma}}\left(\lambda,\rho,\gamma\right)$ is the
probability density function (PDF) of the SINR $\gamma$ observed
at the typical UE for particular values of $\rho$ and $\lambda$.
Based on the definition of $p^{{\rm {cov}}}\left(\lambda,\rho,\gamma\right)$
in (\ref{eq:Coverage_Prob_def}) and the partial integration theorem
shown in~\cite{Renzo2016intensityMatch}, (\ref{eq:ASE_def}) can
be reformulated as%
\begin{comment}
Based on , which is the complementary cumulative distribution function
(CCDF) of SINR, $f_{\mathit{\Gamma}}\left(\lambda,\gamma\right)$
can be computed as 
\begin{equation}
f_{\mathit{\Gamma}}\left(\lambda,\gamma\right)=\frac{\partial\left(1-p^{{\rm {cov}}}\left(\lambda,\gamma\right)\right)}{\partial\gamma}.\label{eq:cond_SINR_PDF}
\end{equation}
\end{comment}
\begin{eqnarray}
A^{{\rm {ASE}}}\left(\lambda,\rho,\gamma_{0}\right)\hspace{-0.1cm}\hspace{-0.1cm} & = & \hspace{-0.1cm}\hspace{-0.1cm}\frac{\tilde{\lambda}}{\ln2}\int_{\gamma_{0}}^{+\infty}\frac{p^{{\rm {cov}}}\left(\lambda,\rho,\gamma\right)}{1+\gamma}d\gamma\nonumber \\
\hspace{-0.1cm}\hspace{-0.1cm} &  & \hspace{-0.1cm}\hspace{-0.1cm}+\tilde{\lambda}\log_{2}\left(1+\gamma_{0}\right)p^{{\rm {cov}}}\left(\lambda,\rho,\gamma_{0}\right).\label{eq:ASE_def_reform}
\end{eqnarray}

Note that $\tilde{\lambda}$ (i.e., the SSR density) is used in the
expression of $A^{{\rm {ASE}}}\left(\lambda,\rho,\gamma_{0}\right)$
because only active BSs make an effective contribution to the ASE,
and that according to (\ref{eq:lambda_tilde_Huang}), $\tilde{\lambda}$
(i.e., the SSR density) is a finite value since $\rho<+\infty$.%
\begin{comment}
The ASE defined in this paper is different from that in~\cite{related_work_Jeff},
where a constant rate based on $\gamma_{0}$ is assumed for the typical
UE, no matter what the actual SINR value is. The definition of the
ASE in (\ref{eq:ASE_def}) can better capture the dependence of the
transmission rate on SINR, but it is less tractable to analyze, as
it requires one more fold of numerical integral compared with~\cite{related_work_Jeff}.
\end{comment}
\begin{comment}
Previously in Subsection~\ref{subsec:The-Coverage-Probability},
we have obtained a conclusion from Theorem~\ref{thm:p_cov_UAS1}:
$p^{\textrm{cov}}\left(\lambda,\gamma\right)$ with the BS IMC should
be better than that with all BSs being active in dense SCNs due to
$\tilde{\lambda}\leq\lambda$. Here from (\ref{eq:ASE_def}), we may
arrive at an opposite conclusion for $A^{{\rm {ASE}}}\left(\lambda,\gamma_{0}\right)$
because it scales linearly with the active BS density $\tilde{\lambda}$,
which caps at $\rho$ in dense SCNs with the BS IMC. The takeaway
message should not be that the IMC generates an inferior ASE in dense
SCNs. Instead, since there is a finite number of active UEs in the
network, some BSs are put to sleep and thus the spatial spectrum reuse
in practice is fundamentally limited by $\rho$. The key advantage
of the BS IMC is that the per-UE performance should increase with
the network densification as discussed in Subsection~\ref{subsec:The-Coverage-Probability}. 
\end{comment}

\subsection{A Constant Capacity Scaling Law\label{subsec:new-cap-scaling}}

\begin{comment}
Placeholder
\end{comment}

From Theorem~\ref{thm:capacity-invariance-law} and the expression
of the ASE in (\ref{eq:ASE_def_reform}), we propose a capacity scaling
law for UDNs in Theorem~\ref{thm:capacity-invariance-law}.
\begin{algorithm*}[t]
\begin{thm}
\label{thm:capacity-invariance-law}A constant capacity scaling law:
If $L>0$ and $\rho<+\infty$, then $\underset{\lambda\rightarrow+\infty}{\lim}A^{{\rm {ASE}}}\left(\lambda,\rho,\gamma_{0}\right)$
becomes a constant that is independent of $\lambda$ in UDNs%
\begin{comment}
, and such constant is determined by $L$, $\rho$ and $\gamma_{0}$
\end{comment}
. In more detail, $\underset{\lambda\rightarrow+\infty}{\lim}A^{{\rm {ASE}}}\left(\lambda,\rho,\gamma_{0}\right)$
is given by
\begin{equation}
\underset{\lambda\rightarrow+\infty}{\lim}A^{{\rm {ASE}}}\left(\lambda,\rho,\gamma_{0}\right)=\frac{\rho}{\ln2}\int_{\gamma_{0}}^{+\infty}\frac{\underset{\lambda\rightarrow+\infty}{\lim}p^{{\rm {cov}}}\left(\lambda,\rho,\gamma\right)}{1+\gamma}d\gamma+\rho\log_{2}\left(1+\gamma_{0}\right)\underset{\lambda\rightarrow+\infty}{\lim}p^{{\rm {cov}}}\left(\lambda,\rho,\gamma_{0}\right),\label{eq:ASE_limit}
\end{equation}

\noindent where $\underset{\lambda\rightarrow+\infty}{\lim}p^{{\rm {cov}}}\left(\lambda,\rho,\gamma\right)$
is obtained from Theorem~\ref{thm:p_cov_limit_UAS1}, and it is independent
of $\lambda$ in UDNs. 
\end{thm}
\begin{IEEEproof}
See Appendix~E.
\end{IEEEproof}
\end{algorithm*}
 The implication of this capacity scaling law in Theorem~\ref{thm:capacity-invariance-law}
is profound, which will be discussed in the following.\vspace{0.1cm}

\textbf{Remark~1:} As discussed in Section~\ref{sec:Introduction},
the conclusion%
\begin{comment}
implication of \emph{the previous SINR invariance} found
\end{comment}
{} in~\cite{Jeff2011} was that the network capacity should scale linearly
as the BS density $\lambda$ increases in a fully-loaded UDN (i.e.,
the SSR density is also $\lambda$). Such conclusion gave us \emph{a}
linear capacity scaling law and showed\textbf{ }an \emph{optimistic}
future for 5G. \vspace{0.1cm}

\textbf{Remark~2:} The implication of Theorem~\ref{thm:capacity-invariance-law}\emph{}%
\begin{comment}
\emph{new capacity scaling law}, which is a constant scaling law,
\end{comment}
{} is quite different. Specifically, the network densification should
be \emph{stopped} at a certain level for a given UE density $\rho$,
because\emph{ }both the coverage probability and the network capacity
will respectively reach a maximum constant value%
\begin{comment}
, leading to a finite frequency reuse factor $\tilde{\lambda}$ defined
by the finite UE density $\rho$
\end{comment}
, thus showing\textbf{ }a \emph{practical} future for 5G. Any network
densification beyond such level of BS density is a waste of both money
and energy. \vspace{0.1cm}

\textbf{Remark~3:} Recently some concerns about network capacity
collapsing in UDNs have emerged, e.g., \emph{the capacity crash} due
to a non-zero BS-to-UE antenna height difference~\cite{Ding2016ASECrash,Atzeni2017ASEcrash},
thus showing\textbf{ }a \emph{pessimistic} future for 5G. However,
it should be noted that such concern was regarding a fully-loaded
UDN. Our results on the \emph{constant} capacity scaling law addresses
this concern. In more detail, even if the UE density is infinite,
the network capacity crash can still be avoided by activating a finite
subset of BSs (i.e., the SSR density is less than $\lambda$) to serve
\emph{a finite subset of UEs} (i.e., the selected UE density is $\rho$)%
\begin{comment}
 in a time division multiple access (TDMA) or frequency division multiple
access (FDMA) manner
\end{comment}
. In other words, instead of letting the network capacity crash with
an aggressive SSR density $\lambda$, our capacity scaling law points
out another approach of dialing the network back to an SSR density
less than $\lambda$, and thus greatly limiting the amount of inter-cell
interference in the network. As a result, the network capacity crash
can be completely avoided.

\vspace{0.1cm}

\textbf{Remark~4:} Following the leads in \textbf{Remark~2}, Theorem~\ref{thm:capacity-invariance-law}
shows that $A^{{\rm {ASE}}}\left(\lambda,\rho,\gamma_{0}\right)$
in (\ref{eq:ASE_def_reform}) reaches $\underset{\lambda\rightarrow+\infty}{\lim}A^{{\rm {ASE}}}\left(\lambda,\rho,\gamma_{0}\right)$
when $\lambda\rightarrow+\infty$. However, achieving such performance
limit might be cost-inefficient due to the investment on the deployment
of BSs as $\lambda\rightarrow+\infty$. Thus, we further propose \textbf{a
BS deployment problem} as follows.\vspace{0.1cm}

\noindent %
\noindent\fbox{\begin{minipage}[t]{1\columnwidth - 2\fboxsep - 2\fboxrule}%
$\quad$For a given UE density $\rho$, there exists an optimal BS
density $\lambda^{*}$ that can achieve a performance result of $A^{{\rm {ASE}}}\left(\lambda,\rho,\gamma_{0}\right)$
that is with a gap of $\epsilon$-percent from $\underset{\lambda\rightarrow+\infty}{\lim}A^{{\rm {ASE}}}\left(\lambda,\rho,\gamma_{0}\right)$,
i.e.,
\begin{eqnarray}
\hspace{-0.1cm}\hspace{-0.1cm}\hspace{-0.1cm}\hspace{-0.1cm}\hspace{-0.1cm}\hspace{-0.1cm}\underset{\lambda}{\textrm{maximize}} &  & \hspace{-0.1cm}\hspace{-0.1cm}\hspace{-0.1cm}\hspace{-0.1cm}\hspace{-0.1cm}1\nonumber \\
\textrm{s.t.} &  & \hspace{-0.1cm}\hspace{-0.1cm}\hspace{-0.1cm}\hspace{-0.1cm}\hspace{-0.1cm}\frac{\left|\underset{\lambda\rightarrow+\infty}{\lim}\hspace{-0.1cm}\hspace{-0.1cm}A^{{\rm {ASE}}}\hspace{-0.1cm}\left(\lambda,\rho,\gamma_{0}\right)\hspace{-0.1cm}-\hspace{-0.1cm}A^{{\rm {ASE}}}\hspace{-0.1cm}\left(\lambda,\rho,\gamma_{0}\right)\right|}{\underset{\lambda\rightarrow+\infty}{\lim}\hspace{-0.1cm}A^{{\rm {ASE}}}\hspace{-0.1cm}\left(\lambda,\rho,\gamma_{0}\right)}=\epsilon.\label{eq:the-deployment-problem}
\end{eqnarray}
\end{minipage}}

\vspace{0.1cm}
Note that the solution $\lambda^{*}$ to the BS deployment problem
(\ref{eq:the-deployment-problem}) would answer the fundamental question
of \emph{``for a given UE density $\rho$, how dense an UDN should
be?''}. It makes sense that such question and answer should depend
on the UE density $\rho$. The intuition is that network densification
should be \emph{stopped} at $\lambda^{*}$, because the network capacity
saturates at $\lambda^{*}$ with a performance gap of $\epsilon$-percent
from $\underset{\lambda\rightarrow+\infty}{\lim}A^{{\rm {ASE}}}\left(\lambda,\rho,\gamma_{0}\right)$.
As shown by (\ref{eq:the-deployment-problem}), the BS deployment
problem solution can be found by numerical search over $A^{{\rm {ASE}}}\hspace{-0.1cm}\left(\lambda,\rho,\gamma_{0}\right)$,
the details of which are omitted here for brevity, but a numerical
example will be shown in the next section.

\vspace{0.1cm}

\textbf{Remark~5:} Following the leads in \textbf{Remark~3}, we
further investigate (\ref{eq:ASE_limit}) and observe that $\underset{\lambda\rightarrow+\infty}{\lim}A^{{\rm {ASE}}}\left(\lambda,\rho,\gamma_{0}\right)$
should be a \emph{concave} function with regard to $\rho$, which
implies an optimal UE density $\rho^{*}$%
\begin{comment}
for the TDMA/FDMA operation among UEs 
\end{comment}
{} that can maximize $\underset{\lambda\rightarrow+\infty}{\lim}A^{{\rm {ASE}}}\left(\lambda,\rho,\gamma_{0}\right)$.
This is because 
\begin{itemize}
\item Lemma~\ref{lem:Pcov_limit_changes_with_rou} states that $\underset{\lambda\rightarrow+\infty}{\lim}p^{\textrm{cov}}\left(\lambda,\rho,\gamma\right)$
decreases as $\rho$ increases,
\item while $\rho$ also linearly scales the terms in (\ref{eq:ASE_limit})
(i.e., the SSR density $\tilde{\lambda}$ converges to $\rho$ in
UDNs due to the limit of one UE per active BS), and 
\item thus, there should exist an optimal UE density $\rho^{*}$ that can
maximize $\underset{\lambda\rightarrow+\infty}{\lim}A^{{\rm {ASE}}}\left(\lambda,\rho,\gamma_{0}\right)$
in (\ref{eq:ASE_limit}), which mitigates the network capacity crash
as discussed in \textbf{Remark~3}.
\end{itemize}
Considering the general expression of the ASE in (\ref{eq:ASE_def_reform}),
we can make such optimization problem more general and propose \textbf{a
UE scheduling problem} as follows.

\vspace{0.1cm}
\noindent\fbox{\begin{minipage}[t]{1\columnwidth - 2\fboxsep - 2\fboxrule}%
$\quad$For a given BS density $\lambda$, there exists an optimal
UE density $\rho^{*}$ that can maximize $A^{{\rm {ASE}}}\left(\lambda,\rho,\gamma_{0}\right)$,
i.e.,
\begin{eqnarray}
\underset{\rho}{\textrm{maximize}} &  & A^{{\rm {ASE}}}\left(\lambda,\rho,\gamma_{0}\right)\nonumber \\
\textrm{s.t.} &  & 0<\rho\leq\lambda.\label{eq:the-scheduling-problem}
\end{eqnarray}
\end{minipage}}

\vspace{0.1cm}
Note that the solution $\rho^{*}$ to the UE scheduling problem (\ref{eq:the-scheduling-problem})
would answer the fundamental question of \emph{``for a given BS density
$\lambda$, what is the optimal user load }$\rho^{*}$\emph{ that
can maximize the ASE?''}. Note that such optimal user load $\rho^{*}$
and the given BS density $\lambda$ implicitly yields \emph{an optimal
SSR density} $\tilde{\lambda}^{*}$ from (\ref{eq:lambda_tilde_Huang}).
Unlike the BS deployment problem (\ref{eq:the-deployment-problem}),
the UE scheduling problem (\ref{eq:the-scheduling-problem}) is more
complicated to solve. Due to the page limit, we will investigate the
solution of (\ref{eq:the-scheduling-problem}) in the journal version
of this work, but a numerical example will be shown in the next section.\textbf{}%
\begin{comment}
\textbf{Remark~3:} Recently some concerns about network capacity
collapsing in UDNs have emerged, e.g., \emph{the capacity crash} due
to a non-zero BS-to-UE antenna height difference~\cite{Ding2016GC_ASECrash},
or a bounded path loss in the near-field (NF) region~\cite{Liu2016NF},
thus showing\textbf{ a pessimistic future for 5G}. However, it should
be noted that the above studies~\cite{Ding2016GC_ASECrash,Liu2016NF}
assumed an infinite UE density in UDNs, which is not realistic. Our
new discovery on the \emph{constant} capacity scaling law addresses
such concerns.
\end{comment}
\textbf{}%
\begin{comment}
\textbf{Remark~3a:} The non-zero BS-to-UE antenna height difference
precludes the existence of the NF effect, since the former one occurs
when the height difference is in the order of meters~\cite{Ding2016GC_ASECrash},
while the latter one emerges when the distance between a transmitter
and a receiver is in the sub-meter region~\cite{Liu2016NF}. With
this in mind, we can see that the NF effect will not occur in practical
UDNs due to the non-zero BS-to-UE antenna height difference, unless
we decide to lower the BS antenna height straight to the UE antenna
height in the future~\cite{Ding2016GC_ASECrash}. But that would
create new problems such as fast shadow fading~\cite{Ding2016GC_ASECrash}
and the optimization of the BS antenna height. 
\end{comment}
\textbf{}%

\section{Simulation and Discussion\label{sec:Simulation-and-Discussion}}

\begin{comment}
Placeholder
\end{comment}

In this section, we present numerical results to validate the accuracy
of our analysis. According to Tables~A.1-3\textasciitilde{}A.1-7
of~\cite{TR36.828}%
\begin{comment}
 and~\cite{SCM_pathloss_model}
\end{comment}
, we adopt the following parameters for the 3GPP Case: $\alpha^{{\rm {L}}}=2.09$,
$\alpha^{{\rm {NL}}}=3.75$, $A^{{\rm {L}}}=10^{-10.38}$, $A^{{\rm {NL}}}=10^{-14.54}$%
\begin{comment}
$BW=10$\ MHz, An important footnote: Note that for the purpose of
numerical calculation, all the distances should be converted to km
because $A^{{\rm {L}}}$ and $A^{{\rm {NL}}}$ are defined for distances
in km. 
\end{comment}
, $P=24$\ dBm, $P_{{\rm {N}}}=-95$\ dBm (with a noise figure of
9\ dB).%
\begin{comment}
Besides, the UE density $\rho$ and the value of $L$ are set to $\left\{ 300,600\right\} \,\textrm{UEs/km}^{2}$
and $\left\{ 8.5,3.5\right\} \,\textrm{m}$.
\end{comment}
\begin{comment}
which leads to $q=4.05$ in (\ref{eq:cov_area_size_PDF_Gamma})~
\end{comment}
\begin{comment}
\begin{tabular}{|l|l|l|}
\hline 
Parameters & Description & Value\tabularnewline
\hline 
\hline 
$d_{1}$ & The cut-off point in the linear LoS probability function & 300$\,$m\tabularnewline
\hline 
$\alpha^{{\rm {L}}}$ & The path loss exponent for LoS & 2.09\tabularnewline
\hline 
$\alpha^{{\rm {NL}}}$ & The path loss exponent for NLoS & 3.75\tabularnewline
\hline 
$A^{{\rm {L}}}$ & The constant value in the path loss function for LoS & $10^{-14.54}$\tabularnewline
\hline 
$A^{{\rm {NL}}}$ & The constant value in the path loss function for NLoS & $10^{-10.38}$\tabularnewline
\hline 
$P$ & The BS transmission power & 24$\,$dBm\tabularnewline
\hline 
$P_{{\rm {N}}}$ & The noise power & -95$\,$dBm\tabularnewline
\hline 
\end{tabular}
\end{comment}
\begin{comment}
Finally, a very wide BS density ranging from $10^{-1}\,\textrm{BSs/km}^{2}$
all the way to $10^{6}\,\textrm{BSs/km}^{2}$ is studied in this section. 
\end{comment}

\subsection{Validation of the Coverage Probability Performance\label{subsec:validation-coverage-probability}}

\begin{comment}
Placeholder
\end{comment}

\begin{figure}[t]
\noindent \begin{centering}
\includegraphics[width=8cm]{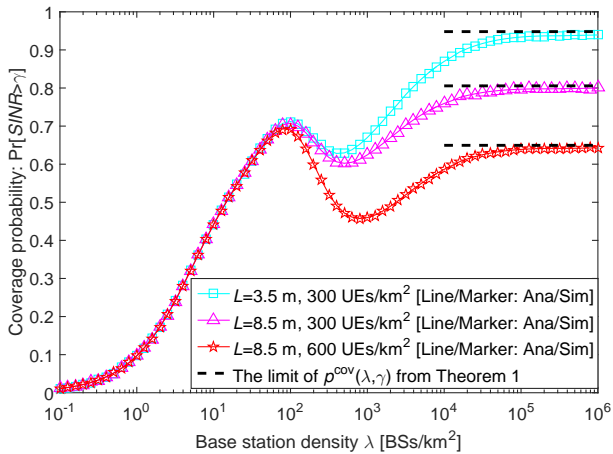}\renewcommand{\figurename}{Fig.}\caption{\label{fig:a-new-SINR-invariance-Pcov-varH-varUEden}The coverage
probability $p^{\textrm{cov}}\left(\lambda,\gamma\right)$ vs. $\lambda$
for the 3GPP Case with $\gamma=0\,\textrm{dB}$ and various values
of $\rho$ and $L$. }
\par\end{centering}
\vspace{-0.3cm}
\end{figure}

In Fig.~\ref{fig:a-new-SINR-invariance-Pcov-varH-varUEden}, we display
the coverage probability for the 3GPP Case with $\gamma=0\,\textrm{dB}$%
\begin{comment}
 and various values of $\rho$ and $L$
\end{comment}
.
 Here, solid lines, markers, and dash lines represent analytical results,
simulation results, and $\underset{\lambda\rightarrow+\infty}{\lim}p^{\textrm{cov}}\left(\lambda,\rho,\gamma\right)$
derived in Theorem~\ref{thm:p_cov_limit_UAS1}, respectively. Note
that the analytical results of $p^{\textrm{cov}}\left(\lambda,\rho,\gamma\right)$%
\begin{comment}
in Fig.~\ref{fig:a-new-SINR-invariance-Pcov-varH-varUEden} 
\end{comment}
{} are obtained from~\cite{Ding2016ASECrash} with $\lambda$ replaced
by $\tilde{\lambda}$.%
\begin{comment}
The details are omitted for brevity. 
\end{comment}
{} From this figure, we can observe that:%
\begin{comment}
\begin{itemize}
\item As already shown in Fig.~\ref{fig:comp_p_cov_4Gto5G}, when the BS
density is around $\lambda\in\left[10^{-1},10^{2}\right]\,\textrm{BSs/km}^{2}$,
the network is noise-limited, and thus $p^{\textrm{cov}}\left(\lambda,\gamma\right)$
increases with $\lambda$ as the network is lightened up with more
BSs and the signal power benefits form LoS transmissions. 
\item As already shown in Fig.~\ref{fig:comp_p_cov_4Gto5G}, when the BS
density is at around $\lambda\in\left[10^{2},10^{3}\right]\,\textrm{BSs/km}^{2}$,
$p^{\textrm{cov}}\left(\lambda,\gamma\right)$ decreases with $\lambda$,
since the network is pushed into the interference-limited region,
and the performance degrades due to the transition of a large number
of interfere paths from NLoS to LoS, which accelerates the growth
of the aggregate inter-cell interference~\cite{related_work_Jeff,our_work_TWC2016}. 
\item When $\lambda\in\left[10^{3},10^{5}\right]\,\textrm{BSs/km}^{2}$,
$p^{\textrm{cov}}\left(\lambda,\gamma\right)$ continuously increases
thanks to the BS idle mode operations~\cite{Ding2016IMC_GC}, i.e.,
the signal power continues increasing with the network densification,
while the interference power becomes bounded, as only BSs with active
UEs are turned on, and thus the number of interfering BSs is controlled
by the number of active UEs.
\end{itemize}
\end{comment}

\begin{itemize}
\item %
\begin{comment}
As already shown in Fig.~\ref{fig:comp_p_cov_4Gto5G}, 
\end{comment}
When the BS density is at around $\lambda\in\left[10^{2},10^{3}\right]\,\textrm{BSs/km}^{2}$,
$p^{\textrm{cov}}\left(\lambda,\rho,\gamma\right)$ decreases with
$\lambda$. This is due to the transition of a large number of interfere
paths from NLoS to LoS, which accelerates the growth of the aggregate
inter-cell interference~\cite{related_work_Jeff,Related_work_Health,our_work_TWC2016}. 
\item When $\lambda\in\left[10^{3},10^{5}\right]\,\textrm{BSs/km}^{2}$,
$p^{\textrm{cov}}\left(\lambda,\rho,\gamma\right)$ continuously increases
thanks to the BS idle mode operations~\cite{Ding2016TWC_IMC}, i.e.,
the signal power continues increasing with the network densification,
while the aggregate interference power becomes bounded, as only BSs
serving active UEs are turned on.
\item When $\lambda>10^{5}\,\textrm{BSs/km}^{2}$, $p^{\textrm{cov}}\left(\lambda,\rho,\gamma\right)$
gradually reaches its limit characterized by Theorem~\ref{thm:p_cov_limit_UAS1},
which verifies the SINR invariance law in Theorem~\ref{thm:SINR-invariance-law}.
Numerically speaking, the gap between the analytical results of $p^{\textrm{cov}}\left(\lambda,\rho,\gamma\right)$
and those of $\underset{\lambda\rightarrow+\infty}{\lim}p^{\textrm{cov}}\left(\lambda,\rho,\gamma\right)$
are less than 0.5$\,$\% for all of the investigated cases when $\lambda=10^{6}\,\textrm{BSs/km}^{2}$,
which validates the accuracy of Theorem~\ref{thm:p_cov_limit_UAS1}. 
\item As shown in Fig.~\ref{fig:a-new-SINR-invariance-Pcov-varH-varUEden},
when $\rho=300\,\textrm{UEs/km}^{2}$, the limit of $p^{\textrm{cov}}\left(\lambda,\rho,\gamma\right)$
with $L=3.5\,\textrm{m}$ is larger than that with $L=8.5\,\textrm{m}$,
thus verifying Lemma~\ref{lem:Pcov_limit_changes_with_L}.
\item As shown in Fig.~\ref{fig:a-new-SINR-invariance-Pcov-varH-varUEden},
when $L=8.5\,\textrm{m}$, the limit of $p^{\textrm{cov}}\left(\lambda,\rho,\gamma\right)$
with $\rho=300\,\textrm{UEs/km}^{2}$ is 0.806, while that with $\rho=600\,\textrm{UEs/km}^{2}$
is 0.65, which equals to the square of 0.806, thus verifying the power
law of $\rho$ in Lemma~\ref{lem:Pcov_limit_changes_with_rou}.
\end{itemize}

\subsection{Validation of the Constant Capacity Scaling Law\label{subsec:validation-new-cap-scaling-law}}

\begin{comment}
Placeholder
\end{comment}

In Fig.~\ref{fig:a-new-cap-scaling-law-ASE-varUEden}, we plot the
ASE results for the 3GPP Case with $\gamma_{0}=0\,\textrm{dB}$, $L=8.5\,\textrm{m}$
and various values of $\rho$. Since $A^{{\rm {ASE}}}\left(\lambda,\rho,\gamma_{0}\right)$
is calculated from the results of $p^{\textrm{cov}}\left(\lambda,\rho,\gamma\right)$
using (\ref{eq:ASE_def_reform}), and because the analysis on $p^{\textrm{cov}}\left(\lambda,\rho,\gamma\right)$
has been validated in Subsection~\ref{subsec:validation-coverage-probability},
we only show the analytical results of $A^{{\rm {ASE}}}\left(\lambda,\rho,\gamma_{0}\right)$
in Fig.~\ref{fig:a-new-cap-scaling-law-ASE-varUEden}. From this
figure, we can observe that:
\begin{itemize}
\item As discussed in \textbf{Remark~1}, due to its simplistic assumptions,
the linear capacity scaling law~\cite{Jeff2011} shows an optimistic
but unrealistic future for 5G UDNs in Fig.~\ref{fig:a-new-cap-scaling-law-ASE-varUEden}.
\item \emph{The constant capacity scaling law} in Theorem~\ref{thm:capacity-invariance-law}
is validated for UDNs with a non-zero $L$ and a finite $\rho$, showing
a practical future for 5G UDNs in Fig.~\ref{fig:a-new-cap-scaling-law-ASE-varUEden},
which has been discussed in \textbf{Remark~2} and \textbf{Remark~3}. 
\item For a given $\rho$, e.g., $\rho=300\,\textrm{UEs/km}^{2}$, the value
of $A^{{\rm {ASE}}}\left(\lambda,\rho,\gamma_{0}\right)$ saturates
as $\lambda\rightarrow+\infty$, which justifies the BS deployment
problem (\ref{eq:the-deployment-problem}) addressed in \textbf{Remark~4}.
For example, for the following set of parameter values: $\rho=300\,\textrm{UEs/km}^{2}$,
$L=8.5\,\textrm{m}$ and $\gamma_{0}=0\,\textrm{dB}$, we can calculate
$\underset{\lambda\rightarrow+\infty}{\lim}\hspace{-0.1cm}A^{{\rm {ASE}}}\hspace{-0.1cm}\left(\lambda,\rho,\gamma_{0}\right)$
using Theorem~\ref{thm:capacity-invariance-law} and obtain its value
as $784.4\,\textrm{bps/Hz/km}^{2}$. Considering a performance gap
of $\epsilon=5$$\,$percent (i.e., a target ASE of $745.2\,\textrm{bps/Hz/km}^{2}$),
it is easy to find the solution to problem (\ref{eq:the-deployment-problem})
as $\lambda^{*}=33420\,\textrm{BSs/km}^{2}$. Such BS density means
that any network densification beyond this level will generate no
more than $5\,\%$ of the maximum ASE. 
\item For a given $\lambda$, e.g., $\lambda=10^{6}\,\textrm{BSs/km}^{2}$,
it is interesting to see that $A^{{\rm {ASE}}}\left(\lambda,\rho,\gamma_{0}\right)$
is indeed a concave function of $\rho$, i.e., $A^{{\rm {ASE}}}\left(\lambda,\rho,\gamma_{0}\right)$
increases when $\rho\in\left[300,600\right]\,\textrm{UEs/km}^{2}$
and decreases when $\rho\in\left[1000,2000\right]\,\textrm{UEs/km}^{2}$.
Hence, it justify the UE scheduling problem (\ref{eq:the-scheduling-problem})
addressed in \textbf{Remark~5}. For example, for the following set
of parameter values: $\lambda=10^{6}\,\textrm{BSs/km}^{2}$, $L=8.5\,\textrm{m}$
and $\gamma_{0}=0\,\textrm{dB}$, we can find the solution to problem
(\ref{eq:the-scheduling-problem}) as $\rho^{*}=804\,\textrm{UEs/km}^{2}$
with a maximum ASE of $928.2\,\textrm{bps/Hz/km}^{2}$. Such optimal
value of $\rho^{*}$ can be translated to an optimal SSR density of
$803.58\,\textrm{SSR/km}^{2}$ from (\ref{eq:lambda_tilde_Huang}).
Note that activating all BSs with a full SSR density of $10^{6}\,\textrm{SSR/km}^{2}$
will lead to the ASE crash~\cite{Ding2016ASECrash,Atzeni2017ASEcrash},
i.e., an ASE of $0\,\textrm{bps/Hz/km}^{2}$.
\item Note that the ASE \emph{crawls} (not increasing quickly) when $\lambda\in\left[10^{2},10^{3}\right]\,\textrm{BSs/km}^{2}$,
which is due to the transition of a large number of interfere paths
from NLoS to LoS~\cite{our_work_TWC2016}.%
\begin{figure}[!t]
\noindent \begin{centering}
\includegraphics[width=8cm]{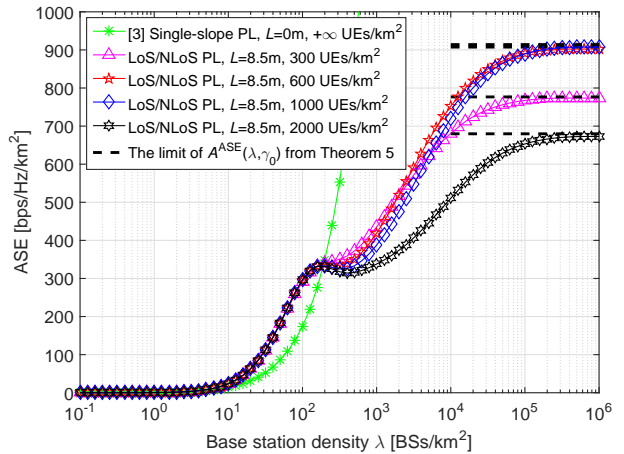}\renewcommand{\figurename}{Fig.}\caption{\label{fig:a-new-cap-scaling-law-ASE-varUEden}The ASE $A^{{\rm {ASE}}}\left(\lambda,\gamma_{0}\right)$
vs. $\lambda$ with $\gamma_{0}=0\,\textrm{dB}$ for the 3GPP Case
with $\gamma_{0}=0\,\textrm{dB}$, $L=8.5\,\textrm{m}$ and various
values of $\rho$. }
\par\end{centering}
\vspace{-0.3cm}
\end{figure}
\end{itemize}

\section{Conclusion\label{sec:Conclusion}}

\begin{comment}
Placeholder
\end{comment}

A constant capacity scaling law has been shown for UDNs. Such law
has two profound implications. First, network densification should
be \emph{stopped} at a certain BS density for a given UE density,
because the network capacity reaches a limit%
\begin{comment}
 due to \emph{(i) }the bounded signal and aggregate interference powers,
and \emph{(ii) }a finite SSR due to the finite UE density
\end{comment}
. Such BS density can be found by solving \emph{the BS deployment
problem} presented in this paper. Second, there exists an optimal
SSR density that can maximize the network capacity. In other words,
when we deploy thousands or millions of BSs per square kilometer,
the best strategy is \emph{not} activating all BSs on the same time/frequency
resource. Such optimal SSR density as well as the corresponding UE
density can be found by solving \emph{the UE scheduling problem} proposed
in this paper. 

\section*{Appendix~A: Proof of Theorem~\ref{thm:p_cov_limit_UAS1}\label{sec:Appendix-A}}

\begin{comment}
Placeholder
\end{comment}

\begin{comment}
Due to the page limit, we only provide the key steps of the proof
as follows. 
\end{comment}
As $\lambda\rightarrow+\infty$, we have that%
\begin{comment}
the 2D distance $r$ from the typical UE to its serving BS $b_{o}$
approaches zero, i.e., 
\end{comment}
{} $r\rightarrow0$ and $w\rightarrow L$ in (\ref{eq:actual_dis_BS2UE}).
Consequently, the path loss of this link should be dominantly characterized
by \emph{the first-piece LoS path loss function} (i.e., $\zeta_{1}^{{\rm {L}}}\left(w\right)$)%
\begin{comment}
, e.g., $L$ is smaller than $d_{1}=67.75$\ m of the 3GPP Case in
practical SCNs~\cite{TR36.828}
\end{comment}
, which supports the use of $\zeta_{1}^{{\rm {L}}}\left(w\right)$
in such case. Thus, $\underset{\lambda\rightarrow+\infty}{\lim}p^{\textrm{cov}}\left(\lambda,\rho,\gamma\right)$
can be derived as
\begin{eqnarray}
\underset{\lambda\rightarrow+\infty}{\lim}p^{\textrm{cov}}\left(\lambda,\rho,\gamma\right)\hspace{-0.1cm}\hspace{-0.1cm}\hspace{-0.1cm} & = & \hspace{-0.1cm}\hspace{-0.1cm}\hspace{-0.1cm}\underset{\lambda\rightarrow+\infty}{\lim}{\rm {Pr}}\hspace{-0.1cm}\left[\mathrm{SINR}\hspace{-0.1cm}>\hspace{-0.1cm}\gamma\left|\zeta\left(w\right)=\zeta_{1}^{{\rm {L}}}\left(L\right)\right.\right]\nonumber \\
\hspace{-0.1cm}\hspace{-0.1cm}\hspace{-0.1cm} & \overset{(a)}{=} & \hspace{-0.1cm}\hspace{-0.1cm}\hspace{-0.1cm}\underset{\lambda\rightarrow+\infty}{\lim}{\rm {Pr}}\hspace{-0.1cm}\left[\frac{P\zeta_{1}^{{\rm {L}}}\left(L\right)h}{I_{{\rm {agg}}}+P_{{\rm {N}}}}\hspace{-0.1cm}>\hspace{-0.1cm}\gamma\right]\nonumber \\
\hspace{-0.1cm}\hspace{-0.1cm}\hspace{-0.1cm} & = & \hspace{-0.1cm}\hspace{-0.1cm}\hspace{-0.1cm}\underset{\lambda\rightarrow+\infty}{\lim}{\rm {Pr}}\hspace{-0.1cm}\left[h\hspace{-0.1cm}>\hspace{-0.1cm}\frac{\left(I_{{\rm {agg}}}+P_{{\rm {N}}}\right)\gamma}{P\zeta_{1}^{{\rm {L}}}\left(L\right)}\right],\label{eq:Proof_1_p_cov_limit}
\end{eqnarray}
where (\ref{eq:SINR}) is plugged into the step (a) of (\ref{eq:Proof_1_p_cov_limit}).
Considering that the complementary cumulative distribution function
(CCDF) of $h$ gives ${\rm {Pr}}\left[h>x_{1}+x_{2}\right]=\exp\left(-x_{1}\right)\exp\left(-x_{2}\right)$
and with some mathematical manipulations, we can arrive at (\ref{eq:Theorem_1_p_cov_limit}).

\begin{comment}
Based on the considered UAS, 
\end{comment}
Then, we can further derive $\mathscr{L}_{I_{{\rm {agg}}}}^{{\rm {L}}}\left(s\right)$
with $s=\frac{\gamma}{P\zeta_{1}^{{\rm {L}}}\left(L\right)}$ as\vspace{0.1cm}

\noindent $\mathscr{L}_{I_{{\rm {agg}}}}^{{\rm {L}}}\left(s\right)$
\begin{eqnarray}
\hspace{-0.1cm}\hspace{-0.1cm} & = & \hspace{-0.1cm}\hspace{-0.1cm}\hspace{-0.1cm}\mathbb{E}_{\left[I_{{\rm {agg}}}\right]}\left\{ \exp\left(-sI_{{\rm {agg}}}\right)\right\} \nonumber \\
\hspace{-0.1cm}\hspace{-0.1cm} & \overset{(a)}{=} & \hspace{-0.1cm}\hspace{-0.1cm}\hspace{-0.1cm}\mathbb{E}_{\left[\tilde{\Phi}\setminus b_{o},\left\{ \beta_{i}\right\} ,\left\{ g_{i}\right\} \right]}\left\{ \exp\left(-s\sum_{i\in\Phi/b_{o}}P\beta_{i}g_{i}\right)\right\} \nonumber \\
\hspace{-0.1cm}\hspace{-0.1cm} & \overset{(b)}{=} & \hspace{-0.1cm}\hspace{-0.1cm}\hspace{-0.1cm}\exp\hspace{-0.1cm}\left(\hspace{-0.1cm}-2\pi\tilde{\lambda}\hspace{-0.1cm}\int_{0}^{\infty}\hspace{-0.1cm}\hspace{-0.1cm}\left(1\hspace{-0.1cm}-\hspace{-0.1cm}\mathbb{E}_{\left[g\right]}\hspace{-0.1cm}\left\{ \exp\hspace{-0.1cm}\left(-sP\beta\hspace{-0.1cm}\left(\hspace{-0.1cm}\sqrt{u^{2}\hspace{-0.1cm}+\hspace{-0.1cm}L^{2}}\right)g\right)\right\} \right)udu\hspace{-0.1cm}\right)\nonumber \\
\hspace{-0.1cm}\hspace{-0.1cm} & \overset{(c)}{=} & \hspace{-0.1cm}\hspace{-0.1cm}\hspace{-0.1cm}\exp\hspace{-0.1cm}\left(\hspace{-0.1cm}-2\pi\tilde{\lambda}\hspace{-0.1cm}\int_{0}^{+\infty}\hspace{-0.1cm}\hspace{-0.1cm}\hspace{-0.1cm}\hspace{-0.1cm}\frac{{\rm {Pr}}^{{\rm {L}}}\left(\sqrt{u^{2}+L^{2}}\right)u}{1\hspace{-0.1cm}+\hspace{-0.1cm}\left(sP\zeta^{{\rm {L}}}\left(\sqrt{u^{2}+L^{2}}\right)\right)^{-1}}du\hspace{-0.1cm}\right)\nonumber \\
\hspace{-0.1cm}\hspace{-0.1cm} &  & \hspace{-0.1cm}\hspace{-0.1cm}\hspace{-0.1cm}\times\exp\hspace{-0.1cm}\left(\hspace{-0.1cm}-2\pi\tilde{\lambda}\hspace{-0.1cm}\int_{0}^{+\infty}\hspace{-0.1cm}\hspace{-0.1cm}\hspace{-0.1cm}\hspace{-0.1cm}\frac{\left[1\hspace{-0.1cm}-\hspace{-0.1cm}{\rm {Pr}}^{{\rm {L}}}\left(\sqrt{u^{2}+L^{2}}\right)\right]u}{1\hspace{-0.1cm}+\hspace{-0.1cm}\left(sP\zeta^{{\rm {NL}}}\left(\sqrt{u^{2}+L^{2}}\right)\right)^{-1}}du\hspace{-0.1cm}\right),\hspace{-0.1cm}\hspace{-0.1cm}\hspace{-0.1cm}\hspace{-0.1cm}\label{eq:laplace_term_LoS_UAS1_seg1_proof_eq1}
\end{eqnarray}
where the step (a) of (\ref{eq:laplace_term_LoS_UAS1_seg1_proof_eq1})
comes from (\ref{eq:cumulative_interference}), the step (b) of (\ref{eq:laplace_term_LoS_UAS1_seg1_proof_eq1})
is obtained from Campbell's theorem~\cite{Jeff2011}, and $\mathbb{E}_{\left[g\right]}\left\{ \exp\left(-sxg\right)\right\} =\frac{1}{1+sx}$
is plugged into the step (c) of (\ref{eq:laplace_term_LoS_UAS1_seg1_proof_eq1})
and the aggregate interference from both LoS and NLoS paths are considered
therein.%
\begin{comment}
\noindent Moreover, in the step (c) of (\ref{eq:laplace_term_LoS_UAS1_seg1_proof_eq1}),
$\textrm{Pr}^{\textrm{L}}\left(r\right)$ is given by 
\begin{equation}
\textrm{Pr}^{\textrm{L}}\left(r\right)=\begin{cases}
\textrm{Pr}_{1}^{\textrm{L}}\left(r\right), & \textrm{when }0\leq r\leq d_{1}\\
\textrm{Pr}_{2}^{\textrm{L}}\left(r\right),\hspace{-0.3cm} & \textrm{when }d_{1}<r\leq d_{2}\\
\vdots & \vdots\\
\textrm{Pr}_{N}^{\textrm{L}}\left(r\right), & \textrm{when }r>d_{N-1}
\end{cases},\label{eq:general_LoS_Pr_proof}
\end{equation}
and $\zeta^{Path}\left(r\right)$ is written as
\begin{equation}
\zeta^{Path}\left(r\right)=\begin{cases}
\zeta_{1}^{Path}\left(r\right), & \textrm{when }0\leq r\leq d_{1}\\
\zeta_{2}^{Path}\left(r\right),\hspace{-0.3cm} & \textrm{when }d_{1}<r\leq d_{2}\\
\vdots & \vdots\\
\zeta_{N}^{Path}\left(r\right), & \textrm{when }r>d_{N-1}
\end{cases},\label{eq:general_PL_func_proof}
\end{equation}
where the string variable $Path$ takes the value of ``L'' and ``NL''
for the LoS and the NLoS cases, respectively. 
\end{comment}
{} Finally, from (\ref{eq:lambda_tilde_Huang}), we have that $\underset{\lambda\rightarrow+\infty}{\lim}\tilde{\lambda}=\rho$,
which yields the result of $\underset{\lambda\rightarrow+\infty}{\lim}\mathscr{L}_{I_{{\rm {agg}}}}^{{\rm {L}}}\left(\frac{\gamma}{P\zeta_{1}^{{\rm {L}}}\left(L\right)}\right)$
in (\ref{eq:laplace_term_LoS_UAS1_general_seg_thm}) and thus concludes
our proof. 

\section*{Appendix~B: Proof of Theorem~\ref{thm:p_cov_limit_UAS1}\label{sec:Appendix-B}}

\begin{comment}
Placeholder
\end{comment}

Due to the page limit, here we only provide the key steps of the proof.
The proof is mainly consisted of two parts, where \emph{(i)} in (\ref{eq:Theorem_1_p_cov_limit}),
$\exp\left(-\frac{P_{{\rm {N}}}\gamma}{P\zeta_{1}^{{\rm {L}}}\left(L\right)}\right)$
is a function of $L$ and $\gamma$. Note that in reality it is a
value very close to 1 because usually we have $P\zeta_{1}^{{\rm {L}}}\left(L\right)\gg P_{{\rm {N}}}$,
and \emph{(ii)} in (\ref{eq:laplace_term_LoS_UAS1_general_seg_thm}),
$\underset{\lambda\rightarrow+\infty}{\lim}\mathscr{L}_{I_{{\rm {agg}}}}^{{\rm {L}}}\left(\frac{\gamma}{P\zeta_{1}^{{\rm {L}}}\left(L\right)}\right)$
is a function of $L$, $\rho$ and $\gamma$. 

\section*{Appendix~C: Proof of Lemma~\ref{lem:Pcov_limit_changes_with_L}\label{sec:Appendix-C}}

\begin{comment}
Placeholder
\end{comment}

Due to the page limit, here we only provide the key steps of the proof.
The proof is mainly consisted of two parts, where for a given $\left\{ \rho,\gamma\right\} $,
\emph{(i)} in (\ref{eq:Theorem_1_p_cov_limit}), we have that $\exp\left(-\frac{P_{{\rm {N}}}\gamma}{P\zeta_{1}^{{\rm {L}}}\left(L\right)}\right)$
decreases as $L$ increases, and \emph{(ii)} in (\ref{eq:laplace_term_LoS_UAS1_general_seg_thm}),
we have that $s=\frac{\gamma}{P\zeta_{1}^{{\rm {L}}}\left(L\right)}$
increases as $L$ increases, and $L$ quickly becomes irrelevant in
the integrals of (\ref{eq:laplace_term_LoS_UAS1_general_seg_thm})
because $L$ appears in the term $\sqrt{u^{2}+L^{2}}$ and the integrals
are performed on $u$ toward $u=+\infty$, which leads to the conclusion
that $\underset{\lambda\rightarrow+\infty}{\lim}\mathscr{L}_{I_{{\rm {agg}}}}^{{\rm {L}}}\hspace{-0.1cm}\left(s\right)$
is a decreasing function of $s$, and thus $L$.%
\begin{comment}
The intuition is that the bounded signal power decreases as $L$ increases. 
\end{comment}
\begin{comment}
Note that Lemma~\ref{lem:Pcov_limit_changes_with_L} supports the
recommendation of lowering the BS antenna height to achieve performance
improvement~\cite{Ding2016GC_ASECrash}.
\end{comment}

\section*{Appendix~D: Proof of Lemma~\ref{lem:Pcov_limit_changes_with_rou}\label{sec:Appendix-D}}

\begin{comment}
Placeholder
\end{comment}

Due to the page limit, here we only provide the key steps of the proof.
First, from (\ref{eq:laplace_term_LoS_UAS1_general_seg_thm}) we conclude
that $\exp\left(-\frac{P_{{\rm {N}}}\gamma}{P\zeta_{1}^{{\rm {L}}}\left(L\right)}\right)$$\approx1$
because usually we have $P\zeta_{1}^{{\rm {L}}}\left(L\right)\gg P_{{\rm {N}}}$
in an interference-limited UDN. Second, the rest of the proof is apparent
from the results in Theorem~\ref{thm:p_cov_limit_UAS1}.%
\begin{comment}
we can see that $\underset{\lambda\rightarrow+\infty}{\lim}\mathscr{L}_{I_{{\rm {agg}}}}^{{\rm {L}}}\left(\frac{\gamma}{P\zeta_{1}^{{\rm {L}}}\left(L\right)}\right)$
ranges from 0 to 1 and it is a decreasing power function of $\rho$.
\end{comment}
\begin{comment}
For example, if we double or triple the value of $\rho$, $\underset{\lambda\rightarrow+\infty}{\lim}p^{\textrm{cov}}\left(\lambda,\gamma\right)$
will decrease by a factor of 4 or 9.
\end{comment}
\begin{comment}
The intuition is that the bounded interference power increases as
$\rho$ increases. 
\end{comment}
\begin{comment}
Note that Lemma~\ref{lem:Pcov_limit_changes_with_rou} supports the
recommendation of employing BS idle modes in 5G~\cite{Ding2016IMC_GC}. 
\end{comment}

\section*{Appendix~E: Proof of Theorem~\ref{thm:capacity-invariance-law}\label{sec:Appendix-E}}

\begin{comment}
Placeholder
\end{comment}

Due to the page limit, here we only provide the key steps of the proof.
As $\lambda\rightarrow+\infty$, the ASE in (\ref{eq:ASE_def_reform})
approaches a limit that is independent of $\lambda$. This is because
\emph{(i)} from Theorem~\ref{thm:p_cov_limit_UAS1}, we can get that
both $\underset{\lambda\rightarrow+\infty}{\lim}p^{\textrm{cov}}\left(\lambda,\rho,\gamma\right)$
and $\underset{\lambda\rightarrow+\infty}{\lim}p^{{\rm {cov}}}\hspace{-0.1cm}\left(\lambda,\rho,\gamma_{0}\right)$
are independent of $\lambda$, and \emph{(ii) }from (\ref{eq:lambda_tilde_Huang})
we have that $\underset{\lambda\rightarrow+\infty}{\lim}\tilde{\lambda}=\rho$,
which is also independent of $\lambda$ and has been plugged into
(\ref{eq:ASE_limit}). Therefore, $\underset{\lambda\rightarrow+\infty}{\lim}\hspace{-0.1cm}A^{{\rm {ASE}}}\left(\lambda,\rho,\gamma_{0}\right)$
is independent of $\lambda$ as $\lambda\rightarrow+\infty$, which
completes our proof. 

\bibliographystyle{IEEEtran}
\bibliography{Ming_library}

\end{document}